\documentclass[aps,prd,showpacs,floatfix,superscriptaddress,
preprintnumbers,nofootinbib,amsmath,amssymb]{revtex4}
\usepackage{graphicx}

\begin{document}

\title{Dynamics of a thin shell in the Reissner--Nordstr\"{o}m metric }

\author{V.~I.~Dokuchaev}
\email{dokuchaev@ms2.inr.ac.ru}
 \affiliation{Institute for Nuclear
Research, Russian Academy of Sciences, pr. 60-letiya Oktyabrya 7a,
Moscow, 117312 Russia}
\author{S.~V.~Chernov}
\email{chernov@lpi.ru}
 \affiliation{Institute for Nuclear
Research, Russian Academy of Sciences, pr. 60-letiya Oktyabrya 7a,
Moscow, 117312 Russia}
 \affiliation{Lebedev Physical Institute,
Russian Academy of Sciences, Leninskii pr. 53, Moscow, 119991
Russia}

\begin{abstract}
We describe the dynamics of a thin spherically symmetric
gravitating shell in the Reissner--Nordstr\"om metric of the
electrically charged black hole. The energy-momentum tensor of
electrically neutral shell is modelled by the perfect fluid with a
polytropic equation of state. The motion of a shell is described
fully analytically in the particular case of the dust equation of
state. We construct the Carter--Penrose diagrams for the global
geometry of the eternal black hole, which illustrate all possible
types of solutions for moving shell. It is shown that for some
specific range of initial parameters there are possible the stable
oscillating motion of the shell transferring it consecutively in
infinite series of internal universes. We demonstrate also that
this oscillating type of motion is possible for an arbitrary
polytropic equation of state on the shell.
\end{abstract}


\maketitle

\section{Introduction}

The model of thin gravitating shells that was first proposed by
Israel \cite{IsraelNC} occupies an important place among the
exactly solvable problems in general relativity. The formalism of
the model of thin shells was subsequently developed in details and
used for a wide class of cosmological and astrophysical problems
\cite{IsraelPR,Kuchar,CruzIsraelPR,BlGuenGuth,BerKuzTkachPRD}. In
particular, when the phase transitions in the early universe are
analyzed \cite{Kirznits,Kirznits2}, the model of thin shells is a
very convenient formalism that allows the dynamics of the phase
transitions themselves and the formation and evolution of baby
universes to be traced in sufficient detail
\cite{SatoSasKodMae1,Kod,Sato,KodSasSatMaed,Sat,KodSas,SasKodSat,KodSasSat,
MaedSasSat,MaedSat,SutSatSat,DolIllarKarNov,BarIsrael,
BerKuzTkach4,BerKuzTkach6,BerKuzTkach5,BerKuzTkach2}. The phase
transitions in the early universe begin with the formation of seed
bubbles of a new vacuum
\cite{ZeldovKobzarOcyn,KobzOkunVoloshin,Col,CalCol,ColLuc}. This
process is a quantum one, but the bubbles of a new vacuum pass
into the classical stage of evolution due to their rapid
expansion. The classical stage of dynamical evolution of vacuum
bubbles was considered using the formalism of thin shells in many
papers \cite{BlGuenGuth,BerKuzTkachPRD,MaedGRG,DokCherPismaJETP,
DokCherCQG,DokCherJETP,BerKuzTkachJETP,AurilDenLegSpalPLB,
AurilDenLegSpalNPB,AurilPalSpal,AguirJohn,KijMagMal}. In
astrophysics, the formalism of thin shells helps to analyze the
relativistic properties of compact stellar systems
\cite{BarBelBisJETP}. In the field theory, models similar to the
model of a thin shell were constructed when the decay dynamics of
a metastable vacuum was studied
\cite{Kirznits,Kirznits2,ZeldovKobzarOcyn,KobzOkunVoloshin,
Col,CalCol,ColLuc,AbbCol,LeeLeeLeePar,LeeLeeLeeNamPar,LeeLeeNamPar}.
The model of thin shells is also convenient for the semiclassical
description of quantum black holes \cite{Berezin,BerezinPRD,
BerBoyNerPRD,BerBoyNerPLB,BerezinArxiv,BerKozKuzTkachPLB,BerezinPLB}.
The special case of a thin shell with a phantom equation of state
falling to a black hole was considered in \cite{BerDokErosSmir}.
The dynamics of a thin rotating dust shell was studied in
\cite{BerOkh}.

The possibility of a stable (oscillatory) motion of a shell in the
metric of an electrically charged Reissner--Nordstr\"{o}m black
hole was recently discussed in \cite{GuenShilCQG}. However, the
corresponding solution was not found. A spherically symmetric
shell in the Minkowski \cite{BerKuzTkachJETP}, Schwarzschild
\cite{IsraelPR,BerKuzTkachPRD}, Schwarzschild–de Sitter
\cite{BlGuenGuth,DokCherCQG} metrics, and the
Friedmann–-Schwarzschild universe \cite{DokCherJETP} is known to
be dynamically unstable. In the long run, the shell either
collapses (falls to the central singularity) or expands
infinitely. In this paper, we investigate in detail the dynamics
of a spherically symmetric shell for the Reissner– geometry and
find the conditions under which the shell oscillations can be
stable. In particular, this problem can be investigated completely
analytically for a dust shell. Kuchar \cite{Kuchar} was among the
first authors who considered the problem on the dynamical
evolution of a shell in the Reissner– metric. In particular, he
showed that the electric charge of the black hole could prevent
the shell collapse, i.e., a bounce point could exist. Previously,
Novikov \cite{Novikov66} showed that the collapse of a charged
sphere could stop and subsequently expand into another universe.
The existence of a bounce point for a contracting shell can
provide its oscillatory motion in the case of an eternal black
hole whose global geometry contains an infinite number of
identical universes. This possibility was discussed qualitatively
in \cite{GuenShilCQG}. Below, we will find necessary conditions
for the realization of such an oscillatory shell motion and the
corresponding exact analytical solution. The same oscillatory
shell motion can be assumed to be possible for the Kerr metric,
because a rotating black hole has a centrifugal barrier. There is
no such possibility for a Schwarzschild black hole and the
collapsing shell inevitably falls to the central singularity.

Everywhere below, we assume that the Greek indices $\alpha$,
$\beta,\ldots$ correspond to four coordinates $t$, $r$, $\theta$
and $\varphi$ in the four dimensional spacetime, while the Latin
indices $i$, $k,\ldots$ correspond to three coordinates $t$,
$\theta$ and $\varphi$ on the shell.

\section{The equations of motion of the shell}

Let there be a spherically symmetric hypersurface $\Sigma$ in the
four dimensional spacetime that divides the spacetime into two
regions. We will denote the inner and outer parts by the
subscripts "in'' and ``ou''. We will describe each region inside
and outside this hypersurface by the metric of an electrically
charged Reissner--Nordstr\"{o}m black hole. It has the well known
form
\begin{eqnarray}
 ds^2_{\rm in,out}&=&\left(1-\frac{2m_{\rm in,out}}{r}
 +\frac{Q_{\rm in,out}^2}{r^2}\right)dt^2
 -\left(1-\frac{2m_{\rm in,out}}{r}
 +\frac{Q_{\rm in,out}^2}{r^2}\right)^{\!-1}\!\!dr^2
-r^2d\Omega,  \nonumber
\end{eqnarray}
where $m_{\rm in,out}$ are the black hole masses and $Q_{\rm
in,out}$ are the electric charges inside and outside the
hypersurface $\Sigma$. Since we will consider an electrically
neutral spherical shell, $Q_{\rm in}=Q_{\rm out}=Q$. By spherical
symmetry, the metric on the shell is \cite{BerKuzTkachPRD}
\begin{eqnarray}
 ds^2_\Sigma=d\tau^2-\rho^2(\tau)d\Omega,
\end{eqnarray}
where $\tau$ is the proper time of an observer located on the
shell and $\rho$ is the shell radius measured by the observer on
the shell. The equations of motion of a thin shell were derived in
many papers (see, e.g., \cite{BerKuzTkachPRD}) and can be written
as
\begin{eqnarray}
 \big[K_0^0\big]+\big[K_2^2\big]=8\pi S_2^2,\quad
 \{K_0^0\}S_0^0+2\{K_2^2\}S_2^2+\big[T_n^n\big]=0,\nonumber\\
  \Big[K_2^2\Big]=4\pi S_0^0,\quad
  \frac{dS_0^0}{d\tau}+\frac{2\dot{\rho}}{\rho}(S_0^0-S_2^2)
 +\big[T_0^n\big]=0,
  \label{GenEquat1}
\end{eqnarray}
where $K_i^j$ is the extrinsic curvature, $T_{\alpha}^{\beta}$ is
the energy– momentum tensor of matter inside and outside the
shell, $S_i^j$ is the energy–momentum tensor of the shell itself,
and the following notation is used: $[T]=T_{\rm out}-T_{\rm in}$
and $\{T\}=T_{\rm out}+T_{\rm in}$. The expressions for the
extrinsic curvature in the Reissner--Nordstr\"{o}m metric are
\cite{Kuchar,BerKuzTkachPRD}
\begin{eqnarray}
\label{K2}
 K_2^2 = -\frac{\sigma}{\rho}\sqrt{\dot{\rho}^2+1-
 \frac{2m}{\rho}+\frac{Q^2}{\rho^2}}, \qquad
 K_0^0 = -\sigma\frac{\ddot{\rho}+m/\rho^2-Q^2/\rho^3}
 {\sqrt{\dot{\rho}^2+1-2m/\rho+Q^2/\rho^2}},
 \label{K0}
\end{eqnarray}
where $\sigma=\pm1$.It can be shown \cite{BerKuzTkach3} that the
signs of $\sigma$ coincide with those of the $R_{+}$ and $R_{-}$
spacetime regions. A charged black hole produces an electric field
outside the black hole whose electromagnetic field tensor is
\cite{Chandra}
\begin{eqnarray}
 F_{tr}=\frac{Q}{r^2}=-F^{tr}.
\end{eqnarray}
The corresponding energy–momentum tensor of the electromagnetic
field is
\begin{eqnarray}
 4\pi T^\alpha_{\,\,\,\beta} = -F^{\alpha\gamma} F_{\beta\gamma}
 +\frac{1}{4}\delta^\alpha_\beta F_{\gamma\delta}F^{\gamma\delta},
 \qquad
 T^t_t = T^r_r=-T^\theta_\theta=-T^\varphi_\varphi
 =\frac{1}{8\pi}\frac{Q^2}{r^4}.
\end{eqnarray}
The corresponding formulas of the passage to the limit
\cite{BerKuzTkachPRD} should be used to calculate the
energy–momentum tensor on a thin shell. As a result, we obtain.
\begin{eqnarray}
 T_0^n=n_{,t}\frac{\partial t}{\partial\tau}T_t^t+
 n_{,r}\frac{\partial r}{\partial\tau}T_r^r=0,\quad
 T_n^n=T_t^t.
\end{eqnarray}
We will consider a shell model with the energy– momentum tensor of
an ideal fluid,
\begin{eqnarray}
 S_{ij}=(p+\mu)u_iu_j-pg_{ij},
\end{eqnarray}
where $p$ is the pressure in the fluid and $\mu$ is its total
energy density. In the reference frame of an observer located on
the three dimensional shell under consideration, the fluid
velocity components are $u_i=(1,0,0)$ For an ideal fluid, the
relations $S_0^0=\mu$ è $S_2^2=S_3^3=-p$ also hold.

Consider a polytropic equation of state $p=kn^\gamma$, where $n$
is the number density and $\gamma$ is the polytropic index. The
total energy density for the polytropic equation of state is
$\mu=n+p/(\gamma-1)$, where the first and second terms correspond
to the fluid particle rest mass and the internal energy density,
respectively \cite{Tooper}. In particular, $p=0$ and $\mu=n$ for
dust; for thermal radiation, the fluid particle rest mass is zero,
$\gamma=4/3$, and $\mu=3p$; for the special case of $\gamma=2$,
$\mu=n+p$. This equation of state for $p\gg n$ reproduces the
ultra-stiff equation of state of a fluid in which the speed of
sound is equal to the speed of light \cite{Zel}: $\mu=p$. For the
polytropic equation of state under consideration, we find the
dependence of the number density $n$ on the shell radius $\rho$
measured by an observer on the shell using the last equation in
(\ref{GenEquat1}):
\begin{eqnarray}
 n=\frac{A}{\rho^2}.
  \label{n}
\end{eqnarray}
Here, A is the constant of integration. Accordingly, the
dependence of the total energy density $S_0^0$ on the shell radius
$\rho$ is
\begin{eqnarray}
 S_0^0=\mu(\rho)=\frac{A}{\rho^2}+
 \frac{k}{\gamma-1}\frac{A^\gamma}{\rho^{2\gamma}}.
  \label{S00}
\end{eqnarray}
As a result, the first equation in (\ref{GenEquat1}), which
describes the shell dynamics, will be written in final form as
\begin{eqnarray}
 \sigma_{\rm in}\sqrt{\dot{\rho}^2+1-\frac{2m_{\rm in}}{\rho}
 +\frac{Q^2}{\rho^2}}-
 \sigma_{\rm out}\sqrt{\dot{\rho}^2+1-\frac{2m_{\rm out}}{\rho}
 +\frac{Q^2}{\rho^2}}
 =4\pi\rho\,\mu(\rho),
 \label{GenEquat2}
\end{eqnarray}
where $\sigma_{\rm in,out}=\pm1$. It is this equation that we will
investigate by assuming that $\mu(\rho)>0$.

\begin{figure}[t]
\includegraphics[width=0.4\textwidth]{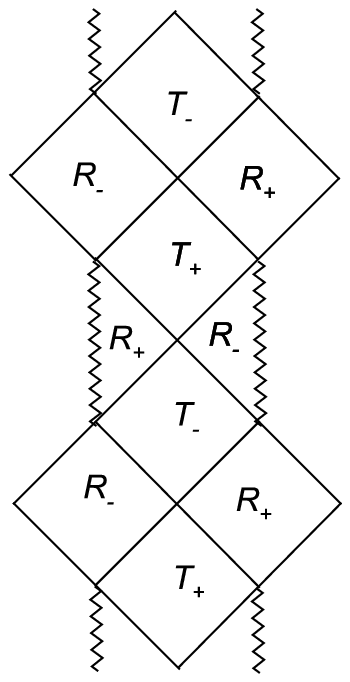}
\caption{The Carter--Penrose diagrams for the metric of an eternal
Reissner--Nordstr\"{o}m black hole (without a shell). The global
geometry of an eternal electrically charged black hole is an
infinite series of identical universes with time-like singularity
(indicated by the polygonal line). The $R_\pm$ and $T_\pm$.
 spacetime regions are shown.} \label{rn}
\end{figure}

For the subsequent analysis, it is convenient to rewrite Eq.
(\ref{GenEquat2}) as the ``energy conservation law''\, by
separating out the kinetic and potential parts. Squaring Eq.
(\ref{GenEquat2}) yields the equation
\begin{equation}
 m_{\rm out}=m_{\rm in}+
 4\pi\rho^2\mu\sigma_{\rm in}\sqrt{\dot{\rho}^2+1-
 \frac{2m_{\rm in}}{\rho}\!+\!\frac{Q^2}{\rho^2}}
 -8\pi^2\rho^3\mu^2.
 \label{energy}
\end{equation}
The quantity $m_{\rm out}$ in this equation is treated as the
Hamiltonian of the entire system \cite{BerKozKuzTkachPLB} (see
also \cite{BerezinPLB}) and is the total energy of the entire
system that is conserved during the dynamical evolution of the
shell. It is easy to assign a physical meaning to other terms in
this equation \cite{BlGuenGuth,BerKuzTkachPRD}. The first term is
the intrinsic mass of the inner black hole. The second term has
the meaning of kinetic energy and the third term is the
gravitational energy of a self-interacting shell with radius
$\rho$. We will call the surface density of this energy an
effective tension of the shell. The Coulomb energy does not enter
into Eq. (\ref{energy}), because the shell considered here is
electrically neutral. Nevertheless, the electric charge of the
black hole enters into the term for the kinetic energy, because it
contributes to the gravitational field of the black hole. Squaring
Eq. (\ref{energy}) once again yields an equation for the shell
evolution in a form convenient for the subsequent analysis:
\begin{equation}
 \dot{\rho}^2+U=0,
 \label{GenEquat3}
\end{equation}
where the effective potential $U=U(\rho)$ is
\begin{equation}
 U=1+\frac{Q^2}{\rho^2}-\frac{m_{\rm out}
 +m_{\rm in}}{\rho}-4\pi^2\rho^2\mu^{2}(\rho)
 -\frac{(m_{\rm out}
 -m_{\rm in})^2}{16\pi^2\rho^4\mu^{2}(\rho)}.
 \label{U}
\end{equation}
The conditions on the signs of $\sigma_{\rm out}$ and $\sigma_{\rm
in}$ that can be obtained using Eq. (\ref{GenEquat2}): should also
be added to this equation:
\begin{eqnarray}
 \label{sigma1}
 \sigma_{\rm out}&=&\mbox{sign}\Bigg[m_{\rm out}-m_{\rm in}
 -8\pi^2\rho^3\,\mu^{2}(\rho)\Bigg], \\
 \sigma_{\rm in}&=&\mbox{sign}\Bigg[m_{\rm out}-m_{\rm in}
 +8\pi^2\rho^3\,\mu^{2}(\rho)\Bigg].
 \label{sigma2}
\end{eqnarray}
In the next section, we investigate these equations in detail for
a dust shell. Similar analysis for other metrics were performed in
many papers \cite{BlGuenGuth,DokCherCQG,DokCherJETP}. Below, using
the Carter--Penrose diagrams, we will construct the global
geometries of the configuration of a charged black hole and a
moving shell under consideration by assuming the black hole to
exist eternally. It is on these diagrams that the entire
evolutionary history of the shell can be traced most completely.
The initial Carter--Penrose diagram for the global geometry of an
eternally existing Reissner--Nordstr\"{o}m black hole without a
shell is shown in Fig. \ref{rn}.

\section{A dust shell}

We will begin our consideration of the shell dynamics with the
simplest case where the shell is a dust one, i.e., the pressure in
it is zero and the energy density is $\mu(\rho)=A/\rho^2$ (see
(\ref{n}) and (\ref{S00})). For a dust shell, the effective
potential $U$ in the equation of motion of the shell
(\ref{GenEquat3}) is simplified considerably and takes the form
\begin{equation}
 U=1+\frac{Q^2}{\rho^2}-\frac{m_{\rm out}+m_{\rm in}}{\rho}
 -\left(\frac{m_{\rm out}-m_{\rm in}}{4\pi A}\right)^{2}
 -4\pi^2\frac{A^2}{\rho^2}.
 \label{Udust}
\end{equation}
The corresponding conditions (\ref{sigma1}) and (\ref{sigma2}) for
the signs of $\sigma$ will be written as
\begin{eqnarray}
 \sigma_{\rm out}=\mbox{sign}\Bigg[m_{\rm out}-m_{\rm in}-
 8\pi^2\frac{A^2}{\rho}\Bigg],\nonumber\\
 \sigma_{\rm in}=\mbox{sign}\Bigg[m_{\rm out}-m_{\rm in}+
 8\pi^2\frac{A^2}{\rho}\Bigg].
\end{eqnarray}
The signs of $\sigma_{\rm out}$ and $\sigma_{\rm in}$ in these
equations change when the shell reaches the radii $\rho_{\rm out}$
and $\rho_{\rm in}$, respectively. These radii are
\begin{eqnarray}
 \rho_{\rm out}=\frac{8\pi^2A^2}{m_{\rm out}-m_{\rm in}},\qquad
 \rho_{\rm in}=\frac{8\pi^2A^2}{m_{\rm in}-m_{\rm out}}.
\end{eqnarray}
It can thus be easily seen that only one of the radii, $\rho_{\rm
out}$, exists at $m_{\rm out}>m_{\rm in}$, while in the opposite
case, i.e., at $m_{\rm in}>m_{\rm out}$, only the second radius
$\rho_{\rm in}$ exists. For the dust equation of state under
consideration, we can find analytically all admissible solutions
to the equation of motion of the shell (\ref{GenEquat3}) and
classify all possible types of shell motion, which we will do
below.

We will begin the classification of all possible types of motion
of a dust shell by studying the characteristic features of the
shape of potential (\ref{Udust}). To find the extrema of this
potential, let us calculate its first and second derivatives:
\begin{eqnarray}
 \frac{\partial U}{\partial\rho}&=&\frac{1}{\rho^2}
 \Bigg[-\frac{2Q^2}{\rho}+m_{\rm in}+m_{\rm out}
 +\frac{8\pi^2 A^2}{\rho}\Bigg], \\
 \frac{\partial^2 U}{\partial\rho^2}&=&\frac{2}{\rho^3}
 \Bigg[\frac{3Q^2}{\rho}-(m_{\rm out}+m_{\rm in})
 -\frac{12\pi^2 A^2}{\rho}\Bigg].
\end{eqnarray}
We see from these equations that the existence or absence of a
potential extremum depends on four characteristic quantities of
our problem: $m_{\rm in}$, $m_{\rm out}$, $Q$ and $A$. The value
of the potential (\ref{Udust}) at spacial infinity $\rho=\infty$
is
\begin{equation}
 U(\infty)=1-\frac{(m_{\rm out}-m_{\rm in})^2}{16\pi^2 A^2}.
\end{equation}
We see that the potential can be both positive and negative,
depending on $m_{\rm out}$, $m_{\rm in}$ and $A$. Accordingly, the
existence or absence of bounce points outside the event horizon of
a black hole will depend on the same parameters. Let us introduce
compact designations for the characteristic outer and inner radii
of the event horizons:
\begin{eqnarray}
\rho_{\rm in}^{\pm}=m_{\rm in}\pm\sqrt{m_{\rm in}^2-Q^2}, \quad
\rho_{\rm out}^{\pm}=m_{\rm out}\pm\sqrt{m_{\rm out}^2-Q^2}.
\end{eqnarray}
For the corresponding potentials at these shell radii, we will
obtain
\begin{eqnarray}
 U(\rho_{\rm in}^{\pm})&=&-\left(\frac{2\pi A}{\rho_{\rm in}^{\pm}}
 +\frac{m_{\rm out}-m_{\rm in}}{4\pi A}\right)^2,\\
 U(\rho_{\rm out}^{\pm})&=&-\left(\frac{2\pi A}{\rho_{\rm out}^{\pm}}
 -\frac{m_{\rm out}-m_{\rm in}}{4\pi A}\right)^2.
 \label{UrhoInOutPM}
\end{eqnarray}
We can also see that the potential on the event horizons is always
negative or zero. The potentials at the characteristic radii
$\rho_{\rm out}$ and $\rho_{\rm in}$ are, respectively,
\begin{eqnarray}
 U(\rho_{\rm out})&=&1-\frac{2m_{\rm out}}{\rho_{\rm out}}
 +\frac{Q^2}{\rho^2_{\rm out}}
 -\left(\frac{2\pi A}{\rho_{\rm out}}
 -\frac{m_{\rm out}-m_{\rm in}}{4\pi A}\right)^2,\nonumber\\
 U(\rho_{\rm in}) & = & 1-\frac{2m_{\rm in}}{\rho_{\rm in}}
 +\frac{Q^2}{\rho^2_{\rm in}}-\left(\frac{2\pi A}{\rho_{\rm in}}
 +\frac{m_{\rm out}-m_{\rm in}}{4\pi A}\right)^2.
 \label{UrhoInOut}
\end{eqnarray}
It follows from these expressions that $\rho_{\rm out}$ and
$\rho_{\rm in}$ coincide with the corresponding event horizons if
the potential on the event horizons and at the characteristic
radii $\rho_{\rm out}$ and $\rho_{\rm in}$ is zero. The potential
is zero, $U=0$, at the shell radius $\rho=\rho_{0}^\pm$, where
\begin{eqnarray}
 \rho_{0}^\pm=\frac{m_{\rm out}+m_{\rm in}}{{2\left[1
 -\frac{(m_{\rm out}-m_{\rm in})^2}{16\pi^2 A^2}\right]}}
 \left\{1\pm\sqrt{1-\frac{4(Q^2\!-\!4\pi^2 A^2)}
 {(m_{\rm out}+m_{\rm in})^2}\left[1-\frac{(m_{\rm out}
 -m_{\rm in})^2}{16\pi^2 A^2}\right]}\right\}.
 \nonumber
\end{eqnarray}
Using the equation for the first derivative of the potential, we
can easily find the radius of its extremum
\begin{eqnarray}
 \rho_{\rm min}=\frac{2(Q^2-4\pi^2A^2)}{m_{\rm out}+m_{\rm in}}.
\end{eqnarray}
This extremum turns out to be a minimum. From the equation for
zero of the second derivative of the potential, we will find the
inflection point
\begin{eqnarray}
 \rho_{\rm inf}=\frac{3(Q^2-4\pi^2A^2)}{m_{\rm out}+m_{\rm in}}
 =\frac{3}{2}\,\rho_{\rm min}.
\end{eqnarray}
At the extremum, the potential is
\begin{eqnarray}
 U(\rho_{\rm min})=1-\frac{(m_{\rm out}-m_{\rm in})^2}{16\pi^2 A^2}
 -\frac{(m_{\rm out}+m_{\rm in})^2}{4\left(Q^2-4\pi^2
 A^2\right)}<0.
 \nonumber
\end{eqnarray}
It can be shown that the potential at the extremum is always
negative.

For a more detailed description, let us consider a situation where
$m_{\rm out}>m_{\rm in}$. The reverse situation is considered
quite similarly. It can be shown that the inequality $\rho_{\rm
out}^+>\rho_{\rm in}^+>\rho_{\rm in}^->\rho_{\rm out}^-$ will hold
at $m_{\rm out}>m_{\rm in}$. This inequality means that the inner
(outer) event horizon of the metric under consideration under the
shell is always larger (smaller) than the corresponding inner
(outer) event horizon of the metric outside the shell. As a
result, the parameter $\sigma_{\rm in}=1$ and the shell moves in
this case in the $R_{+}$ spacetime of the inner part of the
metric. Let us now consider the situation in which $m_{\rm
out}>m_{\rm in}$ for several individual case.

First, let the case (I) where the inequalities $Q^2>4\pi^2 A^2$ è
$(m_{\rm out}-m_{\rm in})^2>16\pi^2 A^2$. hold be realized. The
first inequality means that the branches of the potential $U$ grow
as the shell contracts radially and, hence, the shell will be
unable to overcome the potential barrier. In other words, a bounce
point $\rho_{0}^->0$ exists in this case. The shell cannot fall to
the central singularity, because the gravitational field of the
electric charge of the black hole produces such a high potential
barrier that the kinetic energy of the falling shell is not enough
to overcome it. The second inequality means that the energy of the
shell tension is not enough to prevent the stop of shell
contraction and subsequent expansion to infinity.

Potential (\ref{Udust}) has a minimum at which, as was said above,
it is always negative. Since the potential at infinity is also
negative, the shell expands to an infinite radius in the long run.
It can be shown that the inequality
\begin{equation}
 \rho_{\rm in}^+>\rho_{\rm out}.
\end{equation}
will always hold in this case. This also means that only the
characteristic radii $\rho_{\rm out}^-$, $\rho_{\rm in}^-$ and
$\rho_{\rm out}$ can change places in this case. For a more
detailed description, let us divide this case into two subcases
(Ia and Ib), depending on the relation $\rho_{\rm
out}^{-}\gtrless\rho_{\rm out}$.

\begin{figure}
\includegraphics[width=0.8\textwidth]{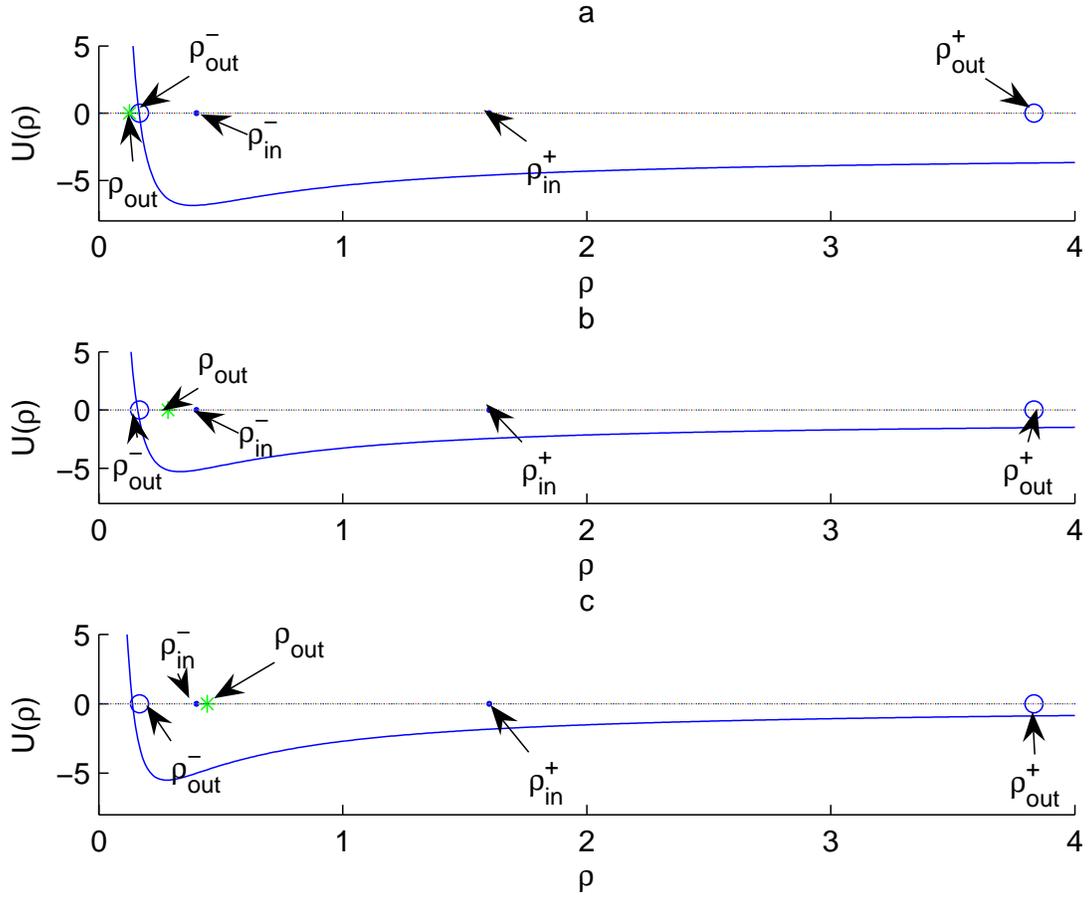}
\caption{Plots of the potentials and positions of the
characteristic radii at $m_{\rm out}=2$, $m_{\rm in}=1$ and
$Q=0.8$ for the following cases: (a) $\rho_{\rm out}^->\rho_{\rm
out}$ at $A=0.04$; (b) $\rho_{\rm in}^->\rho_{\rm out}>\rho_{\rm
out}^-$ at $A=0.06$ and (c) $\rho_{\rm out}>\rho_{\rm in}^-$ at
$A=0.075$. } \label{poten1}
\end{figure}

\begin{figure}
\includegraphics[width=0.4\textwidth]{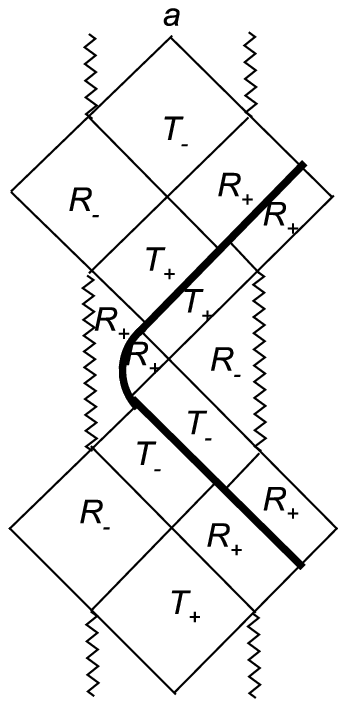}
\includegraphics[width=0.4\textwidth]{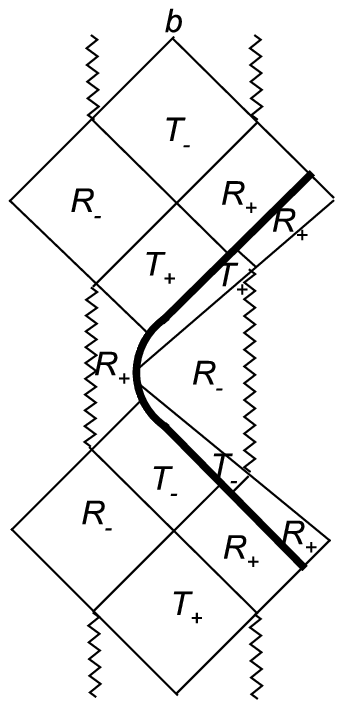}
\caption{Carter--Penrose diagrams for the following cases: (a)
$\rho_{\rm out}^->\rho_{\rm out}$ and  (b) $\rho_{\rm
out}>\rho_{\rm out}^-$. } \label{dust1}
\end{figure}

In the first subcase (Ia) corresponding to the condition
\begin{equation}
\rho_{\rm out}^->\rho_{\rm out},
\end{equation}
or, equivalently, the condition
\begin{eqnarray}
 8\pi^2 A^2<(m_{\rm out}-m_{\rm in})\left(m_{\rm out}-
 \sqrt{m_{\rm out}^2-Q^2}\right),
\end{eqnarray}
the characteristic radii will be located in the following order:
$\rho_{\rm out}<\rho_{\rm out}^-<\rho_{\rm in}^-<\rho_{\rm
in}^+<\rho_{\rm out}^+$, as shown in Fig.~\ref{poten1}a. together
with the plot of $U(\rho)$. The Carter--Penrose diagram
corresponding to this case is presented in Fig.~\ref{dust1}a. On
this diagram, we see that, having begun its contraction from
infinity in the $R_{+}$ region, the shell collapses (i.e., crosses
the outer event horizon of the black hole) and falls into the
$T_{-}$ region. While continuing to contract, the shell then falls
into the inner $R_{+}$ region, where it is reflected from the
potential at the bounce point and begins to expand to infinity
into a new outer $R_{+}$ region, passing through the $T_{+}$
region on its way.

The second subcase (Ib) corresponds to a change in the relative
positions of the radius $\rho_{\rm out}$ and the event horizon
$\rho_{\rm out}^-$. This occurs when the condition $(m_{\rm
out}\!-\!m_{\rm in})\left(m_{\rm out}\!-\!\sqrt{m_{\rm
out}^2\!-\!Q^2}\right)=8\pi^2 A^2$. is met. Thus, for the second
subcase corresponding to the satisfaction of the condition
\begin{eqnarray}
 \rho_{\rm in}^->\rho_{\rm out}>\rho_{\rm out}^-,
\end{eqnarray}
the characteristic radii will be located in the following order:
$\rho_{\rm out}^-<\rho_{\rm out}<\rho_{\rm in}^-<\rho_{\rm
in}^+<\rho_{\rm out}^+$, as shown in Fig.~\ref{poten1}b. The
corresponding Carter--Penrose diagram is presented in
Fig.~\ref{dust1}b. In this case, the shell either expands
infinitely or initially contracts, going below the event horizon,
and subsequently expands to infinity but now in a different
universe.

It remains to consider yet another case (Ic) where

\begin{equation}
\rho_{\rm out}>\rho_{\rm in}^-
\end{equation}
or, equivalently,
\begin{equation}
 (m_{\rm out}-m_{\rm in})\left(m_{\rm in}
 -\sqrt{m_{\rm in}^2-Q^2}\right)<8\pi^2 A^2.
\end{equation}
The characteristic radii will now be located in the following
order: $\rho_{\rm out}^-<\rho_{\rm in}^-<\rho_{\rm out}<\rho_{\rm
in}^+<\rho_{\rm out}^+$, as in Fig.~\ref{poten1}c. The
Carter--Penrose diagram will not change compared to the previous
subcase. As we see from the plots of the potential, there are no
stable, i.e., oscillating solutions for the shell in this case.
The shell always expands to infinity for any initial parameters of
the problem. The general solution for the shell evolution with the
initial condition $\tau=0$ at $\rho=\rho_{0}^-$ that corresponds
to the subcase under consideration will be written as
\begin{eqnarray}
 \tau\sqrt{\!\frac{(m_{\rm out}-m_{\rm in})^2}{16\pi^2 A^2}-\!1}\!
 -\!\sqrt{(\rho-\!\rho_{0}^-)(\rho-\!\rho_{0}^+)}
 \!=\!\frac{\rho_{0}^++\rho_{0}^-}{2}\!
 \ln\!\!\left\{\!\!\!\frac{2\!\!\left[\rho+\!\sqrt{(\rho-\rho_{0}^-)(\rho
 -\rho_{0}^+)}\right]\!\!-\!\rho_{0}^{-}\!-\!\rho_{0}^{+}}{\rho_{0}^{-}
 -\rho_{0}^+}\!\!\right\}\!\!. \nonumber
\end{eqnarray}

Another case (II) is realized when the conditions $4\pi^2 A^2<Q^2$
è $(m_{\rm out}-m_{\rm in})^2<16\pi^2 A^2$ are met.

As will be shown below, when these conditions are met, stable
regular motions of the shell take place. The shell will execute
oscillatory motions. This is because the energy of the shell
tension is enough to reverse the motion of the shell at the bounce
points due to the second inequality. We will also consider all
possible subcases.

\begin{figure}
\includegraphics[width=0.8\textwidth]{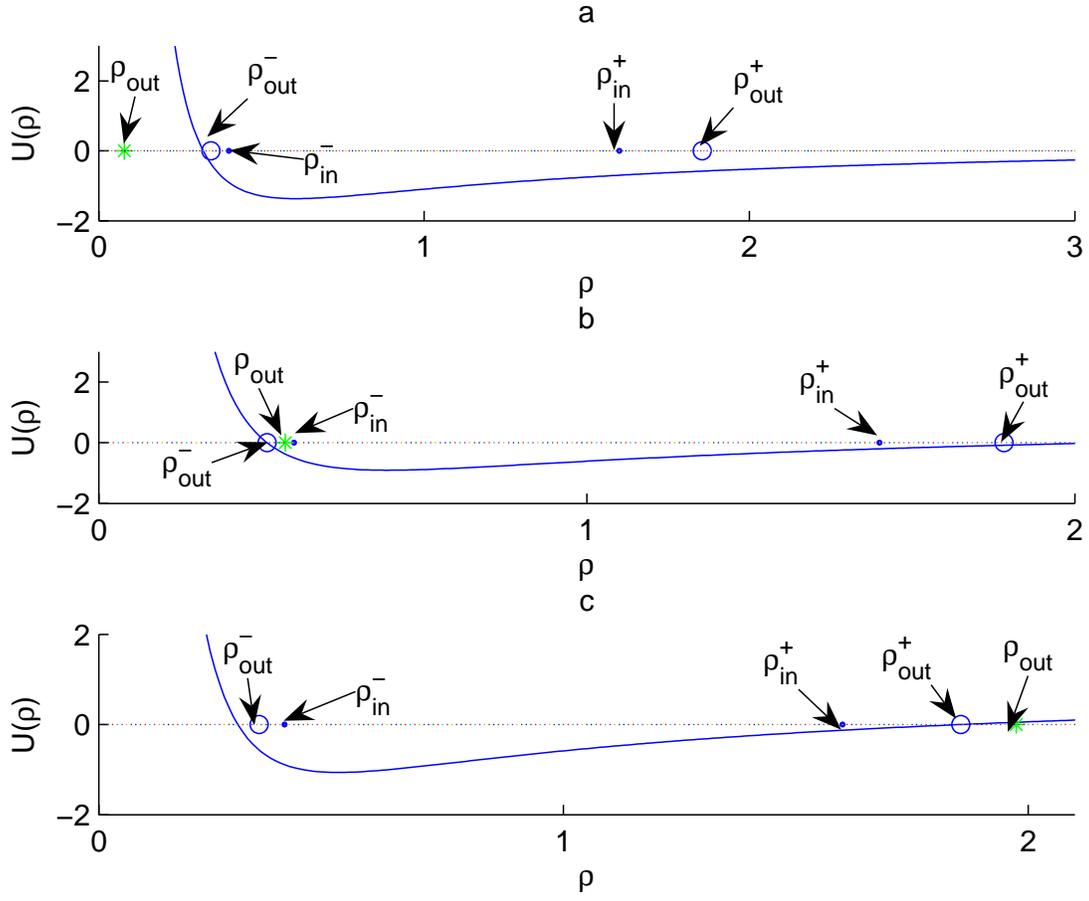}
\caption{Effective potential $U(\rho)$ and relative positions of
the characteristic radii at $m_{\rm out}=1.1$, $m_{\rm in}=1$ and
$Q=0.8$ for the following possible cases: (a) $\rho_{\rm
out}<\rho_{\rm out}^-$ at $A=0.01$; (b) $\rho_{\rm
out}^-<\rho_{\rm out}<\rho_{\rm out}^+$ at $A=0.022$ and (ñ)
$\rho_{\rm out}^+<\rho_{\rm out}$ at $A=0.05$.} \label{poten2}
\end{figure}

\begin{figure}[t]
\includegraphics[width=0.3\textwidth]{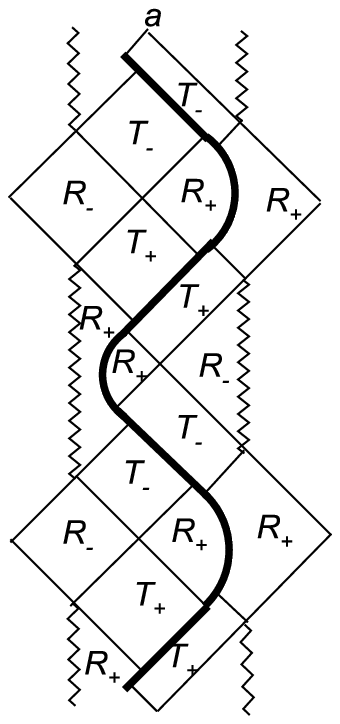}
\includegraphics[width=0.3\textwidth]{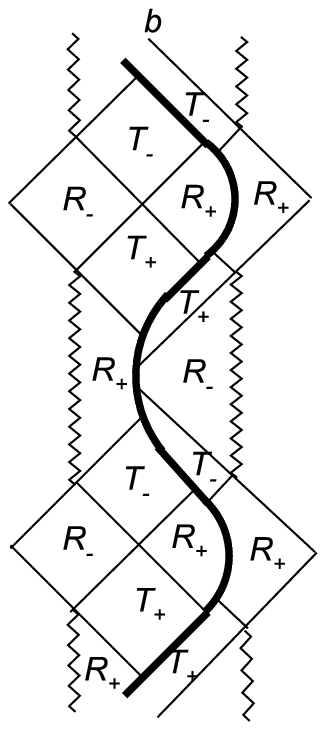}
\includegraphics[width=0.3\textwidth]{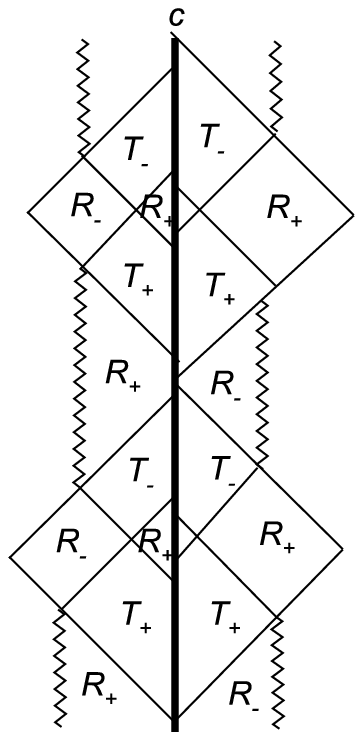}
\caption{The Carter--Penrose diagrams for the following cases: (a)
$\rho_{\rm out}<\rho_{\rm out}^-$; (b) $\rho_{\rm out}^-<\rho_{\rm
out}<\rho_{\rm out}^+$ and (c) $\rho_{\rm out}^+<\rho_{\rm out}$.}
 \label{dust2}
\end{figure}

In the first subcase (IIa), the condition
\begin{equation}
 \rho_{\rm out}<\rho_{\rm out}^-
\end{equation}
is met. The characteristic radii are located in the following
order: $\rho_{\rm out}<\rho_{\rm out}^-<\rho_{\rm in}^-<\rho_{\rm
in}^+<\rho_{\rm out}^+$ as shown Fig.~\ref{poten2}a together with
the plot of the potential. The Carter--Penrose diagrams for this
case are shown in Fig.~\ref{dust2}a.

The next possible subcase (IIb) corresponds to the conditions
\begin{equation}
 \rho_{\rm out}^-<\rho_{\rm out}< \rho_{\rm in}^-.
\end{equation}
The characteristic radii are located in the following order:
$\rho_{\rm out}^-<\rho_{\rm out}<\rho_{\rm in}^-<\rho_{\rm
in}^+<\rho_{\rm out}^+$, as shown in Fig.~\ref{poten2}b. The
Carter--Penrose diagrams is presented in Fig.~\ref{dust2}b.

The next two subcases (IIc and IId) are related to a change in the
relative positions of the characteristic radii $\rho_{\rm out}$
with $\rho_{\rm in}^-$ and $\rho_{\rm in}^+$. In the subcase (IIc)
where
\begin{equation}
 \rho_{\rm in}^-<\rho_{\rm out}<\rho_{\rm in}^+
\end{equation}
the characteristic radii will be located in the following order
$\rho_{\rm out}^-<\rho_{\rm in}^-<\rho_{\rm out}<\rho_{\rm
in}^+<\rho_{\rm out}^+$. Accordingly, in the opposite case where
\begin{equation}
  \rho_{\rm in}^+<\rho_{\rm out}<\rho_{\rm out}^+,
\end{equation}
the characteristic radii will be located in the sequence
$\rho_{\rm out}^-<\rho_{\rm in}^-<\rho_{\rm in}^+<\rho_{\rm
out}<\rho_{\rm out}^+$. The Carter--Penrose diagram remains the
same as that in the previous case.

Finally, in the last possible subcase (IId),
\begin{equation}
 \rho_{\rm out}^+<\rho_{\rm out}.
\end{equation}
In this case, the relative positions of the characteristic radii
are: $\rho_{\rm out}^-<\rho_{\rm in}^-<\rho_{\rm in}^+<\rho_{\rm
out}^+<\rho_{\rm out}$. The potential for this case is plotted in
Fig.~\ref{poten2}c and the Carter--Penrose diagram is shown in
Fig.~\ref{dust2}c. As was said above, two bounce points of the
shell corresponding to radii $\rho_{0}^->0$ and $\rho_{0}^+>0$
exist in this case. We clearly see from the diagrams that the
shell will execute infinite oscillatory motions, successively
passing from one $R_{+}$ region to another $R_{+}$ region. The
corresponding general solution for the evolution of an oscillating
shell with the initial condition $\tau=0$ at $\rho=\rho_{0}^-$
will be written as
\begin{eqnarray}
 \tau\sqrt{1\!-\!\frac{(m_{\rm out}-m_{\rm in})^2}{16\pi^2 A^2}}
 +\!\sqrt{(\rho_{0}^{+}-\rho)(\rho-\rho_{0}^-)}
 \!=\!\frac{\rho_{0}^{+}+\rho_{0}^-}{2}
 \left\{\!\frac{\pi}{2}\!-\!\arctan\!\!\left[\frac{\rho_{0}^
 + +\rho_{0}^{-}-2\rho}{2\sqrt{(\rho_{0}^{+}-\rho)
 (\rho-\rho_{0}^-)}}\right]\right\}. \nonumber
\end{eqnarray}

The next (in order) case (III) arises when the conditions
$4\pi^2A^2>Q^2$ and $(m_{\rm out}-m_{\rm in})^2<16\pi^2 A^2$ are
met. Now, there are no extremum. The branch of the potential grows
smoothly and crosses the axis $U=0$ at point $\rho_{0}^+$

The second bounce point is absent and, accordingly, there is no
oscillating solution. The shell will collapse in the long run.
Even if the shell initially expanded, its expansion will
inevitably change into its contraction, as can be clearly seen on
the Carter--Penrose diagram. It can be shown that the inequality
\begin{equation}
 \rho_{\rm in}^-<\rho_{\rm out}.
\end{equation}
holds in this case. Three additional subcases are possible.

\begin{figure}[b]
\includegraphics[width=0.8\textwidth]{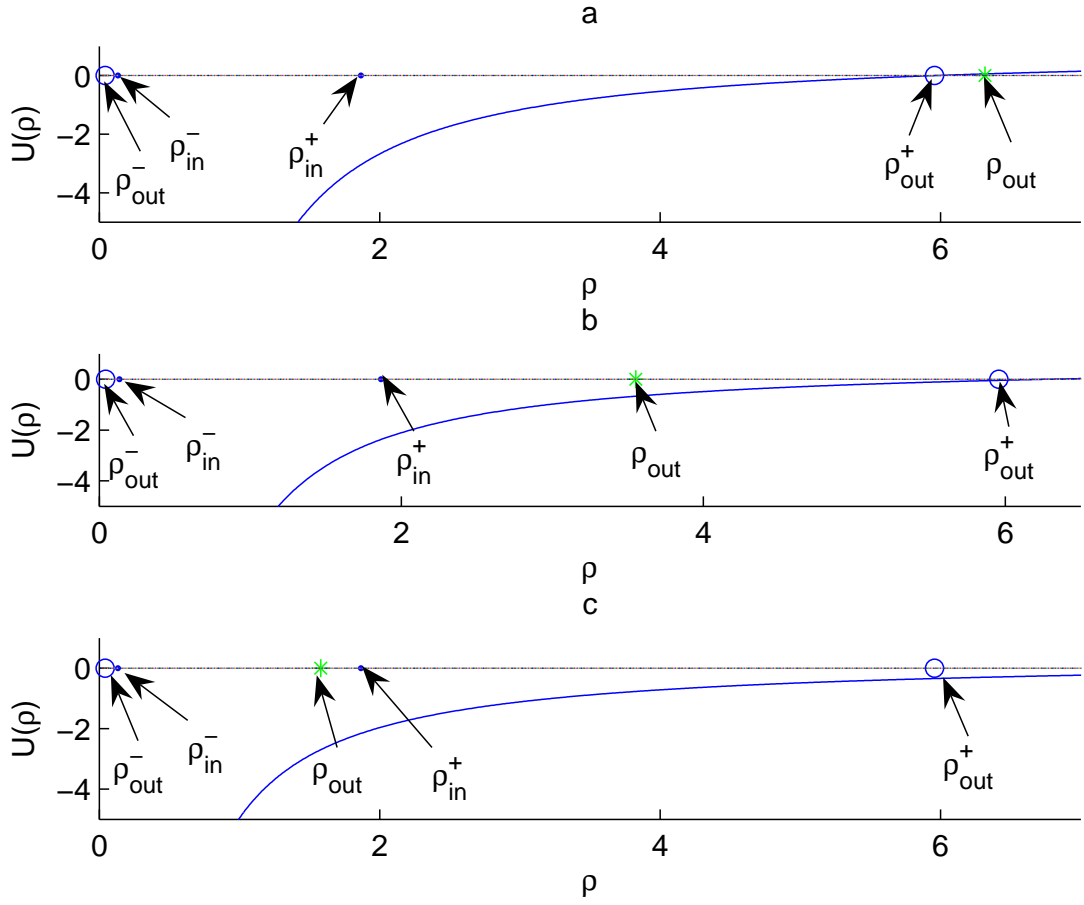}
\caption{Relations $U(\rho)$ and relative positions of the
characteristic radii at $m_{\rm out}=3$, $m_{\rm in}=1$ and
$Q=0.5$ for the following cases:  (a) $\rho_{\rm out}>\rho_{\rm
out}^+$ at $A=0.4$;\, (b) $\rho_{\rm out}^+>\rho_{\rm
out}>\rho_{\rm in}^+$ at $A=0.3$ and (c)
 $\rho_{\rm in}^+>\rho_{\rm out}$ at $A=0.2$.}
 \label{poten3}
\end{figure}

\begin{figure}
\includegraphics[width=0.4\textwidth]{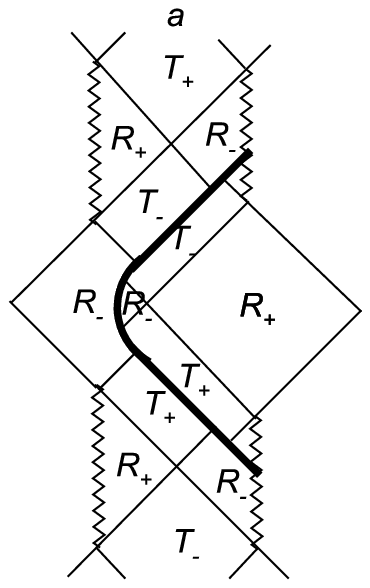}
\includegraphics[width=0.4\textwidth]{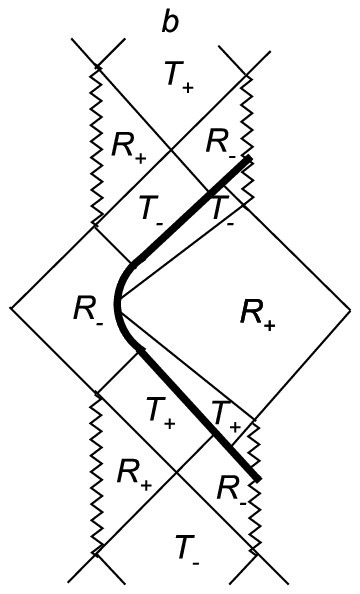}
\caption{The Carter--Penrose diagrams for the following cases: (a)
$\rho_{\rm out}>\rho_{\rm out}^+$ and (b) $\rho_{\rm
out}^+>\rho_{\rm out}$. } \label{dust3}
\end{figure}

In the first subcase (IIIa),
\begin{equation}
 \rho_{\rm out}^+<\rho_{\rm out}.
\end{equation}
The characteristic radii are located in the following order:
$\rho_{\rm out}^-<\rho_{\rm in}^-<\rho_{\rm in}^+<\rho_{\rm
out}^+<\rho_{\rm out}$. The potential for this case is shown in
Fig.~\ref{poten3}a and the Carter--Penrose diagram is presented in
Fig.~\ref{dust3}a.

The second subcase (IIIb) is realized for
\begin{equation}
 \rho_{\rm in}^+<\rho_{\rm out}<\rho_{\rm out}^+.
\end{equation}
This case differs from the previous one only in that the radii
$\rho_{\rm out}$ and $\rho_{\rm out}^+$, change places and, hence,
the relative positions of the characteristic radii are now the
following: $\rho_{\rm out}^-<\rho_{\rm in}^-<\rho_{\rm
in}^+<\rho_{\rm out}<\rho_{\rm out}^+$. The plot of the potential
together with the positions of the characteristic radii is shown
in Fig.~\ref{poten3}b and the Carter--Penrose diagram is presented
in Fig.~\ref{dust3}b.

Finally, the third subcase (IIIc) corresponds to the condition
\begin{equation}
 \rho_{\rm in}^+>\rho_{\rm out}.
\end{equation}
The Carter--Penrose diagram will not change compared to the
previous case. The general solution with the boundary condition
$\tau=0$ at $\rho=\rho_{0}^+$ will be written in this case as
\begin{eqnarray}
 \tau\sqrt{1\!-\!\frac{(m_{\rm out}-m_{\rm in})^2}{16\pi^2 A^2}}
 +\!\sqrt{(\rho_{0}^+-\rho)(\rho-\rho_{0}^-)}\!
 =\!-\frac{\rho_{0}^{+}+\rho_{0}^-}{2}
 \!\left\{\!\!\frac{\pi}{2}\!+\!\arctan\!\!\left[\frac{\rho_{0}^++\rho_{0}^-
 -2\rho}{2\sqrt{(\rho_{0}^+-\rho)
 (\rho-\rho_{0}^-)}}\right]\!\right\}. \nonumber
\end{eqnarray}
Yet another case (IV) is possible where the conditions $4\pi^2
A^2>Q^2$ è $(m_{\rm out}-m_{\rm in})^2>16\pi^2 A^2$ are met.

In this case, there are no bounce points and, hence, the pattern
of shell motion cannot change. Depending on the initial
conditions, the shell either expands infinitely or contracts from
infinity and falls to the singularity.

\section{A polytropic shell}

In this section, we will consider the dynamics of a shell with a
polytropic equation of state that is described by the equation of
motion (\ref{GenEquat3}) with a potential of the general form $U$
(\ref{U}). To analyze the possible types of shell motion, the
equation of motion should be supplemented by the conditions for
the signs of $\sigma$. (\ref{sigma1}) and (\ref{sigma2}). Although
this potential is cumbersome, it has properties similar to those
of the potential for a dust shell. Indeed, for the potential in
the general case, we have the relations:
\begin{eqnarray}
 \label{Uin}
 U(\rho_{\rm in}^{\pm})&=&-\Bigg[2\pi\rho\mu(\rho_{\rm in}^{\pm})
 +\frac{(m_{\rm out}
 -m_{\rm in})}{4\pi\rho_{\rm in}^{\pm2}\mu(\rho_{\rm in}^{\pm})}\Bigg]^2,\\
 U(\rho_{\rm out}^{\pm})&=&-\Bigg[2\pi\rho\mu(\rho_{\rm out}^{\pm})
 -\frac{(m_{\rm out}-m_{\rm in})}
 {4\pi\rho_{\rm out}^{\pm2}\mu(\rho_{\rm out}^{\pm})}\Bigg]^2.
 \label{Uout}
\end{eqnarray}
The values of this potential on the outer and inner event horizons
are negative or zero, as in the dust case. The potential at the
points of change in the sign of $\sigma$ is
\begin{equation}
 U(\rho_{\rm out})=1-
 \frac{2m_{\rm out}}{\rho_{\rm out}}+\frac{Q^2}{\rho_{\rm out}^2}
 - \Bigg[2\pi\rho\mu(\rho_{\rm out})-\frac{(m_{\rm out}
 -m_{\rm in})}{4\pi\rho_{\rm out}^{2}\mu(\rho_{\rm out})}\Bigg]^2,
 \nonumber
\end{equation}
\begin{equation}
 U(\rho_{\rm in}) = 1-\frac{2m_{\rm in}}{\rho_{\rm in}}
 +\frac{Q^2}{\rho_{\rm in}^2}-\Bigg[\!2\pi\rho\mu(\rho_{\rm in})
 +\frac{(m_{\rm out}-m_{\rm in})}
 {4\pi\rho_{\rm in}^{2}\mu(\rho_{\rm in})}\Bigg]^2. \nonumber
\end{equation}
As in the case of a dust shell, it follows from the conditions for
the potential being zero on the event horizon and at the
characteristic radii $\rho_{\rm out}$ and $\rm \rho_{\rm in}$ that
$\rho_{\rm out}$ and $\rm \rho_{\rm in}$ coincide with the
corresponding radii of the black hole event horizons. All of the
above properties of the potential are valid for an arbitrary
equation of state of an ideal fluid. Specifying the equation of
state in an explicit (polytropic) form will be needed only in
calculating the first and second derivatives of the potential:
\begin{eqnarray}
 \frac{\partial U}{\partial\rho}=&-&\frac{2Q^2}{\rho^3}
 +\frac{m_{\rm out}+m_{\rm in}}{\rho^2}-\frac{B(m_{\rm out}
 -m_{\rm in})^2}{4\pi^2\rho^{2\gamma+5}\mu^3(\rho)}
 +8\pi^2\rho\mu(\rho)
 \left[\mu(\rho)+\frac{2B}{\rho^{2\gamma}}\right],
\end{eqnarray}
\begin{eqnarray}
 \frac{\partial^2U}{\partial\rho^2}=&-&\frac{2(m_{\rm out}
 +m_{\rm in})}{\rho^3}+\frac{6Q^2}{\rho^4}
 +8\pi^2\left[\mu(\rho)+\frac{2B}{\rho^{2\gamma}}\right]^2
 -16\pi^2\mu(\rho)\left[\mu(\rho)+\frac{(2\gamma
 +1)B}{\rho^{2\gamma}}\right] \nonumber\\
 &-&\frac{(m_{\rm out}-m_{\rm in})^2(2\gamma-
 1)}{4\pi^2\rho^{2\gamma+6}\mu^3(\rho)}-\frac{3(m_{\rm out}
 -m_{\rm in})^2B^2}{2\pi^2\rho^{4\gamma+6}\mu^4(\rho)},
\end{eqnarray}
where $B=kA^\gamma$. As a result, potential (\ref{U}) for the
equation of motion (\ref{GenEquat3}) of the shell in general form
has the same characteristic features as the potential in the
special case of a dust shell we considered. Consequently, the
parameters of the problem at which an infinite oscillatory motion
of the shell takes place also exist for the general case. For
example, the case of $A=0$ and $\gamma=5/4$ can also be
investigated analytically. However, since the final results are
very cumbersome, we do not present them here. For this case,
stable oscillatory shell motions will take place, in particular,
at the following parameters of the problem: $m_{\rm in}=1$,
$m_{\rm out}=1.01$, $B=0.001$ and $Q=0.5$. Accordingly, for a
different case where $A=0$ and $\gamma=2$, an oscillatory shell
motion will take place, for example, at the following parameters
of the problem: $m_{\rm in}=1$, $m_{\rm out}=1.001$, $B=0.005$ è
$Q=0.9$.

\section{Conclusions}

We analyzed all possible types of dynamical evolution of a thin
shell in the geometry of an eternally existing electrically
charged black hole using a dust shell as an example. In contrast
to the Schwarzschild geometry, in the case of an electrically
charged black hole, apart from the solutions with shell collapse
or infinite expansion, a peculiar oscillating solution
corresponding to the successive shell passage from one universe to
the next in an infinite series of universes inside the event
horizon of the charged black hole also exists. As we see from the
corresponding Carter--Penrose diagrams, a dust shell can travel
infinitely in the inner universes of a charged black hole. In this
case, the periodically alternating stops of shell contraction and
expansion occur successively in different universes. Oscillating
solutions also exist for an arbitrary equation of state of the
shell.

Since the global geometries of a rotating Kerr black hole and an
electrically charged Reissner--Nordstr\"{o}m black hole are
similar, an oscillatory motion of the shell is also possible in
principle in the case of a rotating black hole. Interesting
corollaries of the oscillating solutions can also manifest
themselves in modelling a multiverse that either exists eternally
or results from the evolution of an initially simple universe
through a multiple quantum birth of baby universes and their
subsequent quantum splitting into a set of individual universes
(worlds) connected in a topologically complex way.

Oscillating shells may prove to be the simplest probes that
connect the individual worlds in the multiverse. Indeed, an
observer in our universe can say nothing about the fate of a
collapsing shell. In the case of a Schwarzschild black hole, it
will fall to the central singularity. However, if the black hole
has a charge or an angular momentum, then the shell inside the
black hole can bounce and expand into another universe without
reaching the singularity. If the observer will somehow detect an
expanding shell whose expansion will then change into contraction,
then he can assume that this shell has come from another universe.
Accordingly, electrically charged and rotating black holes with
bounce points inside their event horizons may prove to be the
gates into other universes. Incomplete collapse with a bounce into
another universe opens the fundamental possibility of the
existence of objects from other universes in our universe.

The existence of solutions with a bounce into other universe is of
fundamental importance for the quantization of black holes,
quantum birth of baby universes, and quantum cosmology as a whole,
because the necessity of taking into account the transitions
between states in different universes arises at the quantum level.

Since there is no quantum theory of gravity, semiclassical models
are a powerful tool for investigating quantum effects in strong
gravitational fields. One of the most important present day
problems is the problem of quantum tunnelling in the case of a
multiverse. In particular, the question about the probability of
particle tunnelling from one universe to another through a charged
or rotating black hole arises. The oscillating solutions can be
used to solve the problem of such tunnelling and to consider other
properties of the quantum fields in the Reissner--Nordstr\"{o}m
metric in the semiclassical approximation.

\acknowledgments

This work was supported by the Russian Federal Agency for Science
and Innovation under state contract 02.740.11.5092 and by the
grants of the Leading scientific school 959.2008.2.

\vfill
\newpage

\end{document}